\documentclass{article}
\usepackage[a4paper,margin=2cm,footskip=.5cm]{geometry}
\usepackage{setspace}
\usepackage{graphicx, epstopdf, parskip, latexsym, amsmath, amssymb, mathtools, times, amsthm, url, multicol, rotating, algorithm, algpseudocode, bm}
\usepackage[super,comma]{natbib} 
\usepackage[english]{babel}
\usepackage[utf8]{inputenc}
\usepackage{latexsym}
\usepackage{fancyhdr}
\usepackage{color}
\usepackage{titling}

\theoremstyle{plain}

\theoremstyle{definition}

\theoremstyle{remark}

\DeclareMathOperator{\Cov}{Cov}

\allowdisplaybreaks

\title{Mapping malaria seasonality: a case study from Madagascar} 

\author{Michele Nguyen\thanks{Malaria Atlas Project, Oxford Big Data Institute, Nuffield Department of Medicine, University of Oxford, Oxford, UK. $^{*}$Corresponding author: M. Nguyen (michele.nguyen@bdi.ox.ac.uk). $^{\dagger}$ Joint senior authors.} $^{*}$, Rosalind E. Howes$^{1}$, Tim C.D. Lucas$^{1}$, Katherine E. Battle$^{1}$, \\ Ewan Cameron$^{1}$, Harry S. Gibson$^{1}$, Jennifer Rozier$^{1}$, Suzanne Keddie$^{1}$, Emma Collins$^{1}$, \\ Rohan Arambepola$^{1}$, Su Yun Kang$^{1}$, Chantal Hendriks$^{1}$, Anita Nandi$^{1}$, Susan F. Rumisha$^{1}$,\\ Samir Bhatt\thanks{Department of Infectious Disease Epidemiology, Imperial College London, London, UK.} , Sedera A. Mioramalala\thanks{National Malaria Control Programme, Antananarivo, Madagascar.} , Mauricette Andriamananjara Nambinisoa$^{3}$, \\ Fanjasoa Rakotomanana\thanks{Unité d’Epidémiologie, Institut Pasteur de Madagascar, Antananarivo, Madagascar.} , Peter W. Gething$^{1\dagger}$, Daniel J. Weiss$^{1\dagger}$}

\thanksmarkseries{arabic}

\date{}

\begin{document}

\maketitle
\pagenumbering{arabic}

\thispagestyle{fancy}

\begin{abstract}
Many malaria-endemic areas experience seasonal fluctuations in case incidence as \textit{Anopheles} mosquito and \textit{Plasmodium} parasite life cycles respond to changing environmental conditions. While most existing maps of malaria seasonality use fixed thresholds of rainfall, temperature, and/or vegetation indices to identify suitable transmission months, we develop a statistical modelling framework for characterising the seasonal patterns derived directly from case data.

The procedure involves a spatiotemporal regression model for estimating the monthly proportions of total annual cases and an algorithm to identify operationally relevant characteristics such as the transmission start and peak months. A seasonality index combines the monthly proportion estimates and existing estimates of annual case incidence to provide a summary of “how seasonal” locations are relative to their surroundings. An advancement upon past seasonality mapping endeavours is the presentation of the uncertainty associated with each map, which will enable policymakers to make more statistically sound decisions. The methodology is illustrated using health facility data from Madagascar.
\end{abstract}


\section*{Background} 

Malaria is a disease caused by the \textit{Plasmodium} parasite and remains a major cause of child mortality in sub-Saharan Africa \cite{WHO2018}. As with many vector-borne diseases, malaria transmission exhibits seasonality across endemic areas. That is, malaria burden, which can be measured by metrics including parasite prevalence or the number of clinical cases, follows an annually recurring seasonal pattern that is typically attributed to the relationship of the mosquito vector and parasite life cycles with the environment. The rationale for developing methods capable of enumerating location-specific seasonal characteristics is to assist planning for intervention distributions to improve their efficacy, develop early warning systems, and improve the temporal resolution and overall accuracy of malaria burden estimation models \cite{stuckey2014}.
\\
Past studies on the seasonality of malaria vary in their degree of sophistication and scope. For example, some give empirical descriptions of the cyclic patterns at specific locations, while others model the time series by relating them to underlying seasonal conditions or mathematical structures such as in seasonal autoregressive integrated moving average models and trigonometric models \cite{hamad2002, wardrop2013, dery2010, silal2013, Rumisha2014, wangdi2010}.  
\\
To guide intervention policies, there have also been attempts to derive maps related to seasonality. By using thresholds of, for example, rainfall, temperature, and vegetation cover, it is possible to estimate the start, the end, and the length of the period suitable for transmission \cite{cairns2012, Gemperli2006,  tanser2003}. Maps of malaria seasonality patterns are relevant to the planning of intervention campaigns to maximize their impact. For example, seasonal malaria chemoprevention (SMC) has been shown to be most effective when delivered in areas with highly seasonal transmission. As such, the World Health Organization guidelines recommend targeted SMC in malaria endemic areas where more than 60\% of clinical cases occur during a short period of about four months \cite{WHO2013}. 
\\
Despite their potential utility, the threshold-based malaria seasonality maps have several limitations. For example, metrics such as total rainfall can be linked to either the creation or the washing away of mosquito breeding sites depending on the local topology and rainfall intensity \cite{Barros2011, Valle2014}. Using fixed environmental thresholds does not account for the variation of responses to environmental forcing or allow for other potential drivers such as seasonal migration of human populations \cite{martinez2018, ihantamalala2018}. Likewise, because the distribution of vector species is spatially heterogeneous and their preferences for breeding sites varies, a one-size-fits-all approach for characterizing malaria seasonality may miss important nuances \cite{sinka2016}.
\\
Another class of seasonality maps consists of concentration indices that aim to quantify the proportion of an annual amount (of any variable of interest) which falls within a sub-annual window of fixed size. In order to quantify the distribution of malaria cases in each district over a year, Mabaso et al. used Markham’s concentration index, a measure previously used to determine rainfall concentrations \cite{mabaso2005}. In their analysis, the concentration maps from the case numbers that were estimated using a Bayesian spatiotemporal regression model displayed clearer spatial patterns compared to those derived from raw case numbers. Spatiotemporal models smooth out idiosyncratic deviations, thus enabling the main seasonal trend to be discerned more easily. They are also useful for explicitly relating the seasonality to input covariates as well as accounting for unknown spatiotemporal effects.
\\
In this paper, we improve upon previous attempts to map malaria seasonality and present a modelling framework for a cohesive and evidence-based analysis. Health facility data is used to establish location-specific intra-annual distributions of cases. These inform a log-linear spatiotemporal regression model so that monthly case proportions can be estimated across the study region. From these estimates and fits to rescaled von Mises densities, we identify seasonality features such as the peak and length of transmission. By applying the algorithm to posterior samples of the monthly proportion curves, we obtain uncertainty measures associated with each derived seasonal characteristic. The focus on the monthly case proportions is motivated by a seasonality index which synthesises these estimates with existing annual case or parasite incidence (API) estimates to reflect both the timing and the amplitude aspects of seasonality. The methodology is applied to 2013-2016 health facility case data from Madagascar, a country of marked ecological and epidemiological diversity that is struggling to meet targets for malaria burden reductions and hence where further information for targeting interventions would be valuable \cite{WHO2018, Kang2018}.

\section*{Results}

\subsection*{Dominant environmental relationship}

Malaria seasonality in Madagascar was modelled using monthly case reports submitted to the centralised Ministry of Health between 2013-2016 from 2669 health facilities. Median case counts for each month were derived for each location. This accounted for year-on-year trends and helped avoid unwanted bias from, for example, stock-outs of rapid diagnostic tests (RDTs). Monthly proportions were obtained by dividing the monthly case medians by their annual sum. This is illustrated for an example Malagasy health centre in Supplementary Fig. 1. 
\\
By modelling proportions instead of the case counts, as was done by Mabaso et al., we bypass the need to estimate catchment populations for the health facilities. Additional features of this modelling approach include: (a) utilising a standardised definition of a transmission period as the months with the proportion of cases exceeding $1/12$ of the annual total; and (b) restricting the analysis to dynamic environmental covariates like rainfall and temperature that are likely to impact seasonal malaria transmission patterns.
\\
A log-linear spatiotemporal regression model was fitted to the empirical monthly case proportions. The regression component allows us to infer the dominant relationship between the monthly case proportions and the environmental covariates while the spatiotemporal random field accounts for deviations from this. As expected, the selected model (Table \ref{tab1}) suggests positive relations between the monthly case proportions and rainfall for the concurrent month as well as at two- and three-month lags \cite{reiner2015, weiss2015, nanvyat2018}. There is also a positive relation with the \textit{Plasmodium vivax} temperature suitability index at a two-month lag. The latter is an indicator of the number of infected mosquitoes which are supported by the environment based on the temperature \cite{gething2011, weiss2014}. Since this was derived from a biological model, it accounts for a non-linear relationship with the monthly proportions. 

\subsection*{Seasonality categories and monthly parasite incidence estimates}

\begin{table}[tbp]
\centering
\begin{tabular}{l l l l}
\textbf{Description} & \textbf{Term} & \textbf{Posterior median} & \textbf{95\% credible interval} \\
\hline \\
Intercept & Intercept & $\mathbf{-2.655}$ & $(-2.740, -2.570)$ \\
Precipitation & CHIRPS\_r & $\mathbf{0.060}$ & $(0.029, 0.090)$ \\ 
& CHIRPS\_r\_lag1 & 0.019 & $(-0.012, 0.050)$ \\ 
& CHIRPS\_r\_lag2 & $\mathbf{0.051}$ & $(0.021, 0.082)$ \\ 
& CHIRPS\_r\_lag3 & $\mathbf{0.053}$ & $(0.022, 0.084)$ \\ 
Temperature suitability & TSI\_Pv\_r & 0.013 & $(-0.022, 0.047)$ \\ 
& TSI\_Pv\_r\_lag2 & $\mathbf{0.074}$ & $(0.039, 0.109)$ \\ 
Vegetation cover & EVI\_r\_lag3 & 0.006 & $(-0.021, 0.032)$ \\
Tasselled cap brightness & TCB\_r & -0.019 & $(-0.050, 0.012)$ \\ 
& TCB\_r\_lag3 & 0.011 & $( -0.020, 0.043)$ \\ 
Observation variance & $\sigma_{e}^{2}$ & $0.326$ & $(0.318, 0.333)$ \\
Field variance & $\sigma_{f}^{2}$ &  $0.245$ & $(0.221, 0.268)$ \\
Mat\'ern scaling parameter & $\kappa$  & $3.163$ &  $(2.834, 3.522)$ \\
Autoregressive parameter & $a$ & $\mathbf{0.756}$ & $(0.718, 0.777)$ 
\end{tabular}
\caption{Parameter posterior summaries of the refitted model for Madagascar. The posterior medians of the statistically significant parameters under a 5\% significance level are highlighted in bold. In R-INLA, the Mat\'ern smoothness parameter $\nu$ is typically fixed to $1$ for two spatial dimensions.}\label{tab1}
\end{table}

\begin{figure}[tbp]
\centering
\includegraphics[width = 6.8in, height = 4.2in]{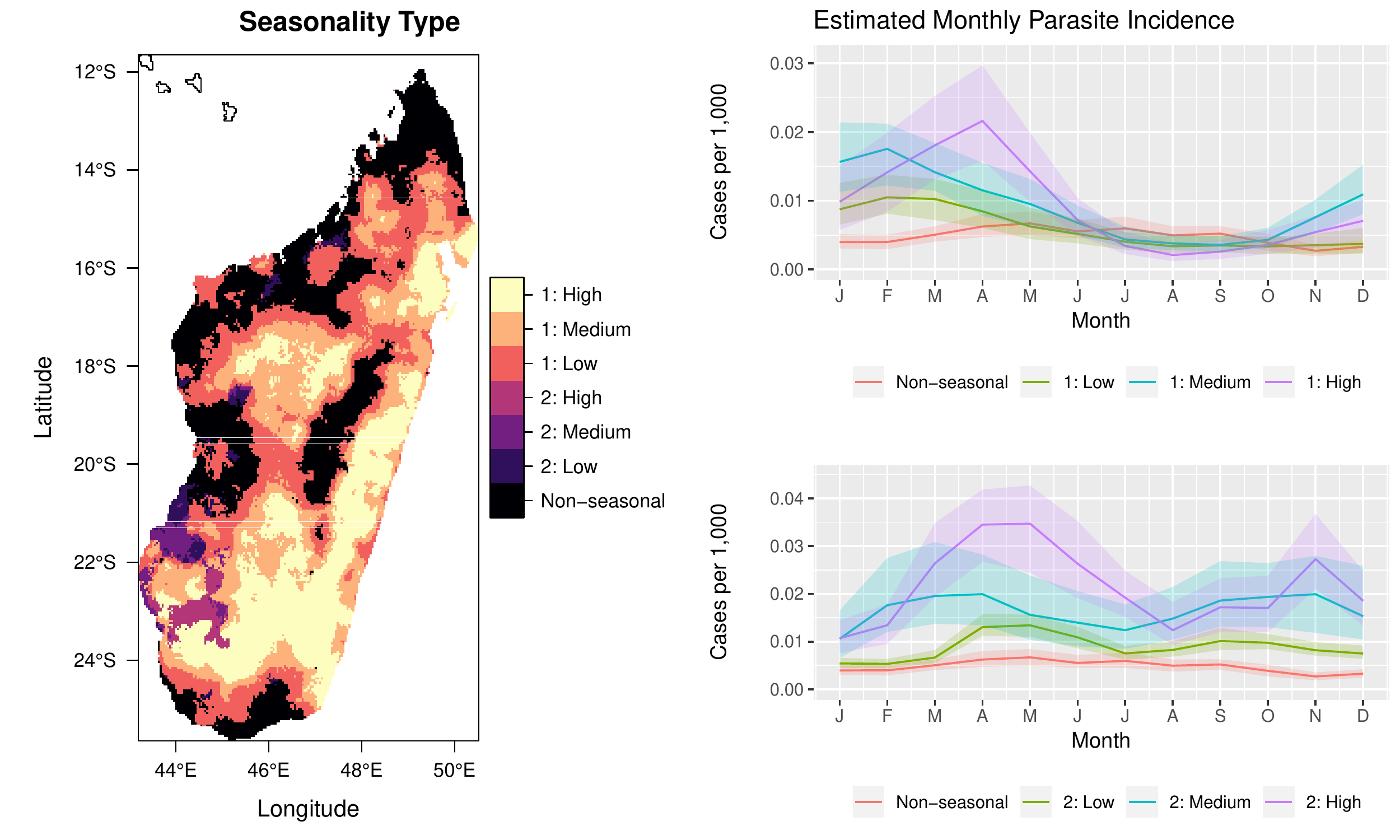}
\caption{Map of seasonality types based on quartiles of the estimated seasonality index as well as representative examples of the estimated monthly parasite incidence for the categories. Here, ``1'' and ``2'' refers to unimodal and bimodal intra-annual distributions respectively while ``Low'', ``Medium'' and ``High'' refers to the different degrees of seasonality.}\label{fig:seas_type_ts}
\end{figure}

``How seasonal'' malaria is in a location is related to both the magnitude and the intra-annual distribution of cases. This can be quantified using a seasonality index. Figure \ref{fig:seas_type_ts} shows the map of seasonal types derived from the median seasonality index, as computed using 100 realisations from the fitted model for Madagascar. The different degrees of seasonality (``Non-seasonal”, ``Low”, ``Medium” and ``High”) were defined using the quartiles of the seasonality indices for pixels with unimodal and bimodal intra-annual distribution of cases modelled separately. In the algorithm, a location is deemed as unimodal or bimodal if more than half of the posterior samples indicate so. 
\\
Examples of the estimated monthly parasite incidence (MPI) curves for each seasonal category are shown alongside the map. In general, we observe that higher seasonality indices are associated with higher average levels of MPI as well as greater amplitudes of the fluctuations. As shown in Supplementary Fig. 2, the bimodal locations (i.e. those with two seasonal peaks in MPI) tend to have lower seasonal index values than the unimodal locations because their distributions are more spread out over the year.

\subsection*{Seasonal features and associated uncertainties}

\begin{figure}[tbp]
\centering
\includegraphics[width = 3.2in, height = 3.2in]{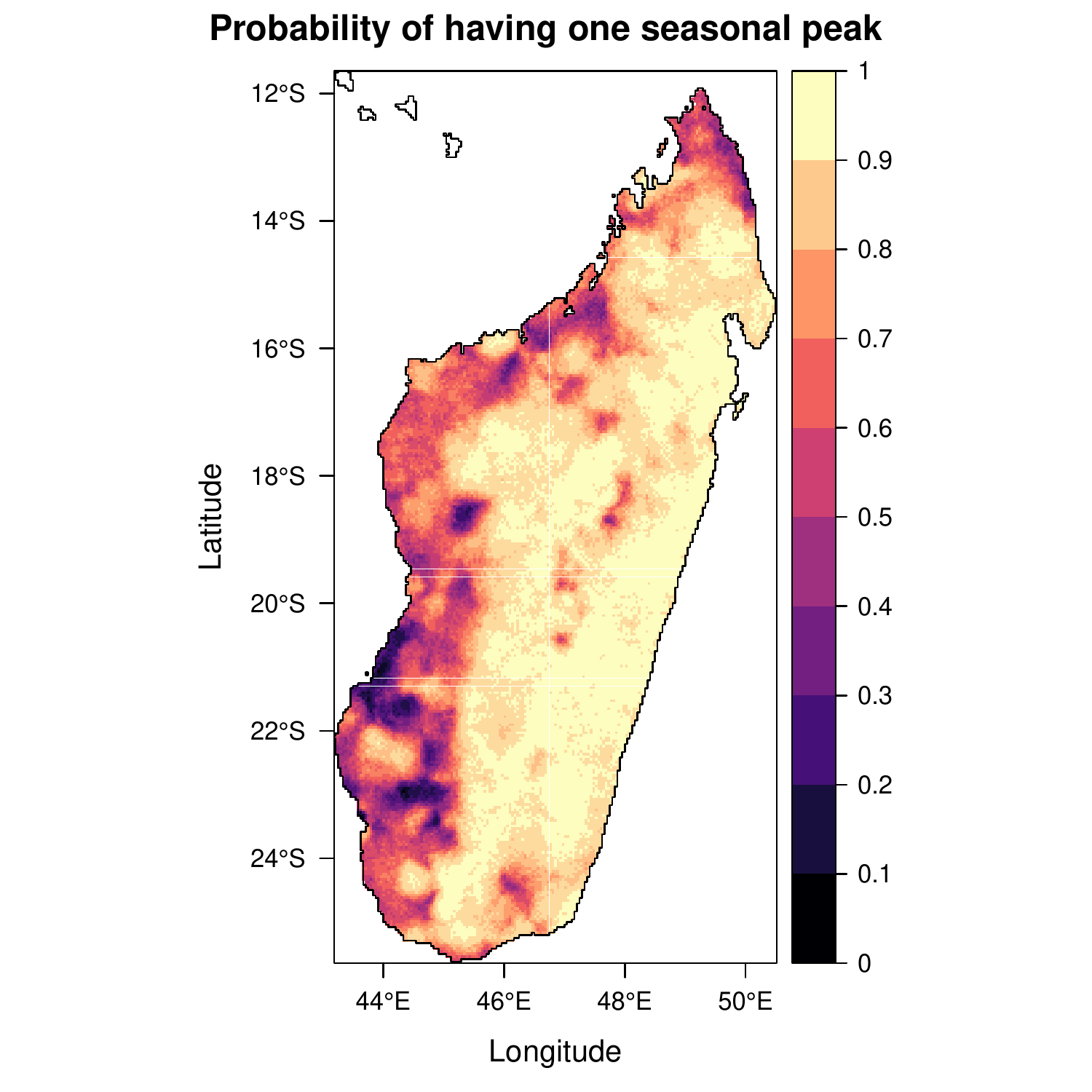}
\caption{Probability of locations having one seasonal peak in malaria cases. This is calculated by the proportion of posterior samples which indicate that the locations have unimodal intra-annual case distributions rather than bimodal distributions.}\label{mad_modality}
\end{figure}

\begin{figure}[tbp]
\centering
\includegraphics[width = 6.4in, height = 3.2in]{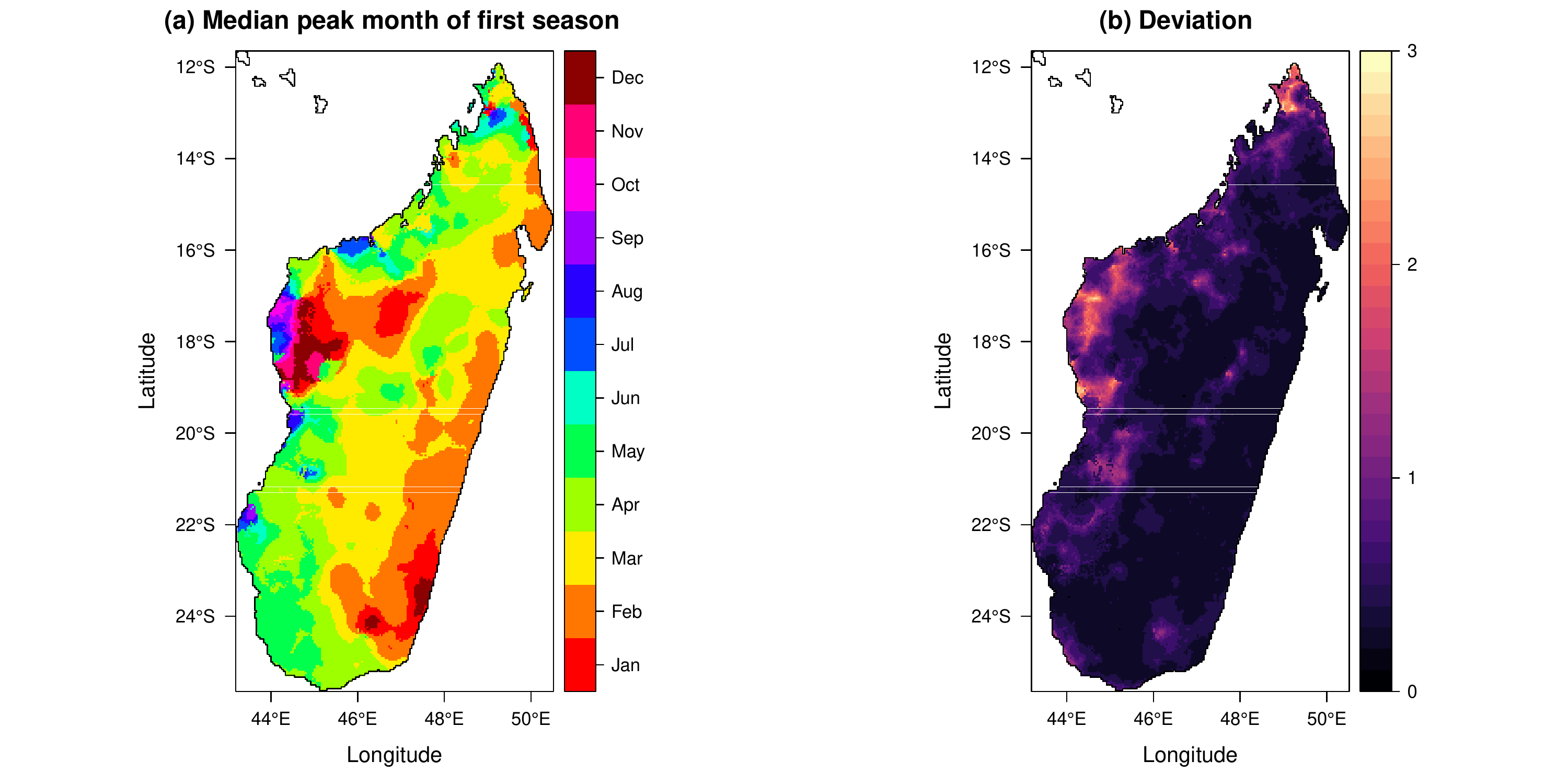}
\caption{(a) Median peak months of the first transmission season in Madagascar and (b) the associated deviations.}\label{mad_peak1}
\end{figure}

\begin{figure}[tbp]
\centering
\includegraphics[width = 6.4in, height = 3.2in]{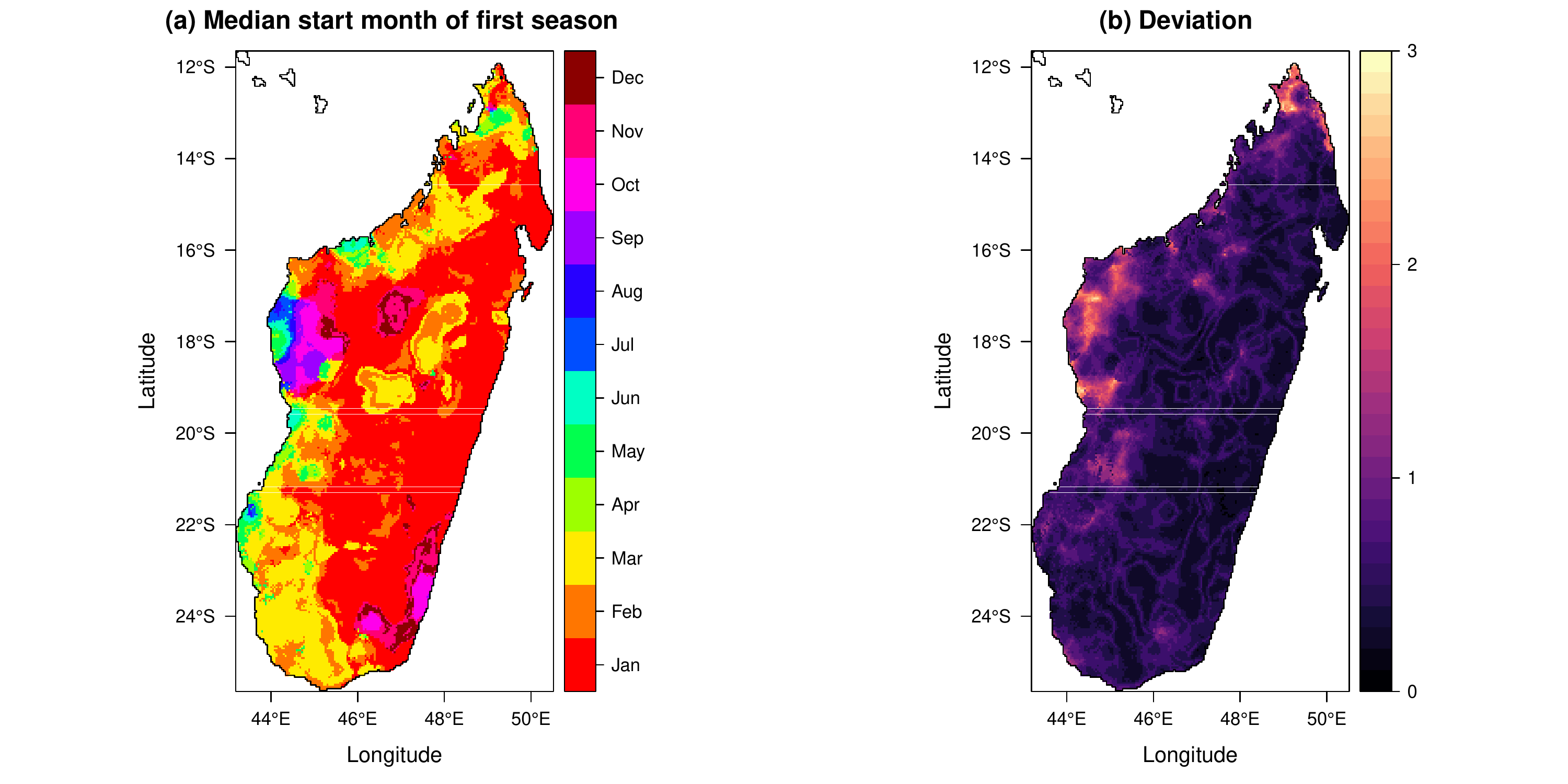}
\caption{(a) Median start months of the first transmission season in Madagascar and (b) the associated deviations.}\label{mad_start1}
\end{figure}

Next, we focus on the timing aspect of seasonality and derive seasonality characteristics such as the start and peak months of transmission. Since we work with the estimated monthly case proportions which rely on the environmental covariates and spatiotemporal correlation, we also obtain results for areas deemed `non-seasonal' via the seasonal index in Fig \ref{fig:seas_type_ts}. Following the definition of the seasonality index in equation (\ref{eqn:seas_ind}) of the ``Methods'' section, such non-seasonality could arise due to a relatively uniform intra-annual distribution of cases or extremely low malaria burden. In the latter scenario, the derived seasonality features describe a theoretical transmission season which could materialise if transmission re-establishes itself. 
\\
From Fig. \ref{fig:seas_type_ts}, we see that most of the island was deemed to have unimodal seasonality. However, as shown in Fig. \ref{mad_modality}, there is generally less certainty along the western and northern coasts. This may be linked to lower data availability (see Supplementary Fig. 5) as well as the remoteness of these regions \cite{howes2016}: lower reporting rates and differing care-seeking behaviour, which could arise due to the lower accessibility of the health facilities, can cause conflicting seasonal signals in the data. Furthermore, distinct rainfall patterns have been described based on the analysis of Liemann et al. which found an increased likelihood of bimodal rainfall patterns on the west coast and unimodal trends on the east coast \cite{liebmann2012}.
\\
The median peak months and associated deviations of the first transmission season are shown in Fig. \ref{mad_peak1}. The results are consistent with existing literature \cite{howes2016}. Large parts of the island experience peaks near March-April while the east coast sees an earlier peak around February. The heterogeneity in Fig. \ref{mad_peak1}(a) in the western region of Melaky (near $45^{\circ}E, 17^{\circ}S$) is associated with high deviations. This is likely due to its remoteness and low population density\cite{howes2016}. In Supplementary Fig. 3, we show the time series of the number of people tested positive for malaria via RDTs between 2013-2016 at three example health facilities in Melaky. The relatively low and highly variable case numbers lead to higher stochasticity in the observed and estimated seasonality patterns. Reporting difficulties, as illustrated by the multiple gaps in the time series, add further uncertainty to the derived monthly proportion curves.
\\
Indoor residual spraying (IRS) is included in the 2018-2022 National Strategic Plan as a tool to help reduce transmission in the highest disease burden districts and as an emergency response tool to epidemic outbreaks \cite{nmcp2017, pmi2019}. Given that sprayed insecticide generally remains efficacious for less than six months (depending on the insecticide used and surface types sprayed)\cite{dengela2018}, local seasonality patterns are important to guide optimal timing of IRS campaigns. Spraying must be timed for completion ahead of the start of transmission, but not so early that the insecticide bio-efficacy will wear off before the end of the season. Figure \ref{mad_start1} shows the median start months and associated deviations of the first transmission season. The results presented here can therefore help guide the planning, with IRS in south-east districts needing to be completed several months ahead of the southwest, for example. To incorporate the higher uncertainty in the model predictions for the southwest, a buffer period of about one month can be used for this region. Additional seasonality plots can be found in the Supplementary Information.

\section*{Discussion}

This paper introduces a statistical modelling framework for mapping malaria seasonality. The approach relies on a log-linear spatiotemporal regression model to smooth and estimate location-specific monthly proportions of cases. Health facility data from Madagascar were used to illustrate the methodology. As countries increasingly adopt digital surveillance platforms such as the District Health Information Software 2 (DHIS2) for the recording of cases at the health facility level, it is hoped that more of such data will be available to inform seasonal and localised intervention strategies.
\\
The modelling framework gives cohesive results and leverages existing API maps to bring together the amplitude as well as timing aspects of seasonality. Characteristics such as the start, peak, and length of each transmission season as well as MPI estimates can be obtained via the same estimated curve of monthly proportions. The latter can also be used to compute a seasonality index and categorise seasonality types. For each seasonal feature, measures of uncertainty can be presented to facilitate statistically sound decision-making.
\\
Despite the many advantages of this approach, there are some limitations. An important assumption was made when the empirical monthly proportions were computed by averaging case counts over a number of years. The notion that there should be a static seasonal trend over multiple years is a common and practical one; however, while climatic patterns are broadly predictable there is significant inter-annual variability in factors such as the beginning of the rainy season. The impact of the annual cyclone season in Madagascar is particularly significant in this respect, triggering both unusually high rainfall (and subsequently increased mosquito vector abundance) and infrastructure damage which can severely disrupt malaria control intervention efforts, resulting in unusual patterns of malaria outbreaks \cite{howes2016, worldbank2017}. Likewise, periodic events related to El Niño-Southern Oscillation and/or global climate change also impact malaria seasonal patterns \cite{reiner2015}. Given this reality, future iterations of this work could use a moving window approach to continually update the model with new data. 

\clearpage

\clearpage

\begin{figure}[tbp]
\centering
\includegraphics[width = 6.4in, height = 2in]{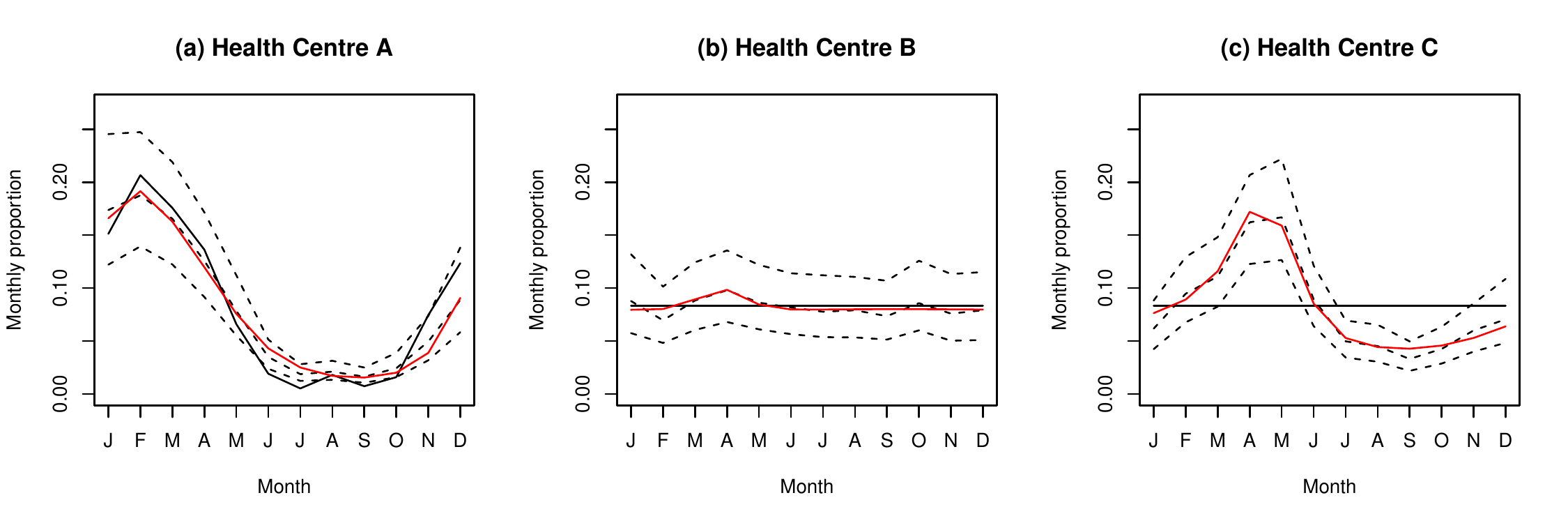}
\caption{Examples of the model fit and rescaled von Mises density fit for three health facilities. The black line denotes the empirical monthly proportions of cases, the black dotted lines represent the median proportions and 95\% credible intervals, and the red line the fitted rescaled von Mises density. Note that no cases were reported for Health Centres B and C, leading to a uniform empirical intra-annual distribution.}\label{mad_validationeg}
\end{figure}

Another noteworthy limitation relates to relying on sparse, spatially discrete health facility reports to establish seasonal patterns across space. Although the issue of errors due to RDT stock-outs was mitigated somewhat by averaging case counts over several years, the issue of zero recorded cases in a facility still poses a potential problem as zero reported cases could be due to the low parasite prevalence (whether historical or only recent) or the low popularity of the facility itself. Another consideration is whether a zero is a true representation of cases in a facility rather than a place-holder for unreported data. While we have tried to ensure the accuracy of the response data by, for example, omitting health facilities with insufficient data to establish a full year-long seasonal trend, it is possible that flawed points were included in the model. This possibility is illustrated in Fig. \ref{mad_validationeg}, which shows the empirical and fitted monthly proportions for three health facilities. Health Centre B (Fig. \ref{mad_validationeg}b) illustrates a scenario where there were no cases and the model estimated a non-seasonal trend. In contrast, Health Centre C (Fig. \ref{mad_validationeg}c) had no cases but an estimated seasonal trend. The clearest interpretation of a pattern in the absence of data is that, while there were no cases reported, the environmental conditions in that area suggest that if there were to be any cases they would tend to peak in April. Such an interpretation is analogous to the theoretical peak transmission season in a country that has eliminated malaria, yet continues to experience seasonally high vector densities. 
\\
The presented model establishes the dominant relationship between environmental covariates and monthly proportions over the study area. This is driven by data in obviously seasonal locations such as Health Centre A in Fig. \ref{mad_validationeg}a. The spatiotemporal field accounts for other unknown factors and smooths out the estimated monthly proportions between locations. In this way, information is borrowed from the seasonal locations identified within the data and applied to areas with similar environmental profiles.
\\
In addition to the level of malaria burden, spatial scale affects the amount of stochasticity in seasonality analyses. Although we bypassed the issue of catchment populations by modelling monthly proportions instead of case numbers, the number of cases seen at a health facility will be more variable if it serves less people. This was seen for the Melaky region in the Madagascar case study. If we aggregate the cases to the administrative (district) level, it is likely that a stronger seasonal signal will be observed. The trade-off is that the relation between administrative level seasonality and area-representative environmental covariates (e.g. average rainfall) may be less strong. 
\\
The seasonality we model is limited by the nature of our data and the available seasonal signal as well as the relations we can establish. Since we work with case data, if treatment seeking behaviour itself is seasonal and related to environmental factors such as rainfall, the seasonality we observe and hence model is merely the seasonality of cases at health facilities which may not be reflective of the seasonal trends for cases at the population level.
\\
Our model assumes that there is one main dominant relation between case numbers and the environmental factors. Although the spatiotemporal field helps adjust for differences, different settings can give rise to different responses to environmental forcing. For example, one frequently links increased mosquito breeding habitats to the period after the rainy season. However, in the Brazilian Amazon this is linked to the dry season when small, isolated water bodies are created with the receding of rivers. While local knowledge may be used to split up the study regions based on differing responses, more research is required on the different settings and how to integrate them into our model structure.
\\
Malaria seasonality maps are useful for targeting interventions such as seasonal malaria chemoprevention and indoor residual spraying. Even with the limitations outlined above, the proposed modelling framework represents the state-of-the-art in obtaining evidence-based seasonality maps and estimates. With the ability to infer the dominant environmental relation in the study region as well as to give uncertainty measures with each seasonal feature, it presents an advancement from the existing threshold and concentration-based mapping procedures. It also provides cohesive results for a seasonality index, monthly case incidence estimates and transmission season statistics. As more health facility case data becomes available, such analyses will become more pertinent. 

\section*{Methods}

\subsection*{Spatiotemporal monthly proportion model} \label{sec:stmodel}

To estimate monthly case proportions over the study region, the following spatiotemporal model was used:
\begin{equation}
\log(p_{i,j}) = \bm{X}_{ij}\bm{\beta} + \phi_{ij} + \epsilon_{ij}, \label{eqn:propmodel}
\end{equation}
where $p_{i,j}$ is the average proportion of cases at location $j$ in month $i$, $\bm{X}_{ij}\in \mathbb{R}^{n\times m}$ is a covariate design matrix including an intercept and $\bm{\beta}\in \mathbb{R}^{m}$ is the corresponding parameter vector. The spatiotemporal Gaussian field $\phi$ is constructed such that:
\begin{equation}
\phi_{i,j} = \begin{cases} \xi_{1j} \text{ for } i = 1, \\
 a \phi_{i-1,j} + \xi_{i, j} \text{ for } i=2, \dots, 12,
\end{cases}
\label{eqn:st.field}
\end{equation}
$|a|<1$ and $\xi_{i, j}$ correspond to zero-mean Gaussian innovations which are temporally independent but spatially coloured with a Mat\'ern covariance:
\begin{equation}
\Cov(h) = \frac{\sigma^{2}_{f}}{\Gamma(\nu)2^{\nu-1}}(\kappa h)^{\nu} K_{\nu}(\kappa h), \label{eqn:Matern}
\end{equation}
where $h$ is the distance between two locations and $\kappa>0$ is a scaling parameter. 
\\
Model fitting and selection were performed in R using integrated nested Laplace approximation (INLA) \cite{Kang2018, Rue2009, Lindgren2011, Martins2013}. The order of the modified Bessel function of second kind, denoted by $K_{\nu}$ in equation (\ref{eqn:Matern}), was set to $1$, the INLA default value. In addition, by specifying $\epsilon \sim N(0, \sigma_{e}^{2})$ in equation (\ref{eqn:propmodel}) to be independent, identically distributed noise, we assume a Gaussian likelihood for our monthly case proportions. 
\\
Note that to avoid applying logarithms on zeros, we added an offset of $0.00001$ to $p_{i, j}$ and rescaled the raw monthly proportions at each location to sum to one before modelling. Unlike a spatiotemporal logistic regression or a compositional regression, the ratios from the model (\ref{eqn:propmodel}) were not modelled with respect to a fixed reference month. Thus, it was easier to relate each month’s proportion to the values of its covariates explicitly. Rescaling of the estimated proportions is required to ensure that they sum to one at each location. This is consistent with the fact that some locations are more or less sensitive to the variation in the underlying factors. 
\\
Before fitting the model, the log-proportions were examined and outliers were excluded to model prototypical seasonal behaviour. Based on the histogram of $\log(p_{i,j})$ values in Supplementary Fig. 4, monthly proportions with $\log(p_{i,j})\leq -11$ were deemed as outliers in the Madagascar analysis. To standardize the covariates, monthly medians of annual covariate data that were temporally aligned with the health facility data were also derived. The covariates consisted of the Climate Hazards Group Infrared Precipitation and Station data (CHIRPS), enhanced vegetation index (EVI), daytime land surface temperature (LST\_day), diurnal difference in land surface temperature (LST\_delta), night-time land surface temperature (LST\_night), tasselled cap brightness (TCB), tasselled cap wetness (TCW), the temperature suitability indices for \textit{Plasmodium falciparum} and \textit{Plasmodium vivax} (TSI\_Pf and TSI\_Pv). Details on the sources of these data are available elsewhere \cite{Kang2018, weiss2015}. To account for delayed and accumulated responses to these environmental variables, 1-3 month lags were also included for each covariate.
\\
To validate the model, 30\% of sites, randomly selected, were excluded from the model fitting. Working with data from the remaining 70\% of the locations, the set of covariates was reduced to facilitate model selection and account for multicollinearity by iteratively computing the variance inflation factors (VIFs) and removing the covariates with the highest VIF value until all the remaining covariates have VIF values less than 10. Since these covariates have low correlations with each other, their estimated regression coefficients are more robust.  
\\
To speed up the backwards regression for the Madagascar analysis, the 70\% training set was randomly split into two smaller sets of equal size and spatial coverage to search for the best model in terms of Deviance Information Criterion (DIC). A map of the test and training locations is shown in Supplementary Fig. 5. The two resulting candidates from the separate backwards regressions on the two smaller training sets were then refitted to the whole training set to select the final model with the lower DIC. After checking for reasonable results on the test data, this parsimonious model was refitted to the entire data set before predictions were made over the gridded surface. 

\subsection*{Seasonality index and monthly case incidence}

Seasonality statistics for each posterior sample of the location-specific monthly proportions were derived after refitting the chosen model to the complete dataset. To quantify ``how seasonal'' a location is, a seasonality index was defined \cite[]{feng2013}. For location $j$, this index is given by the product of an entropy measure and the normalised amplitude:
\begin{align}
S_{j} &= D_{j} \times \frac{R_{j}}{R_{\text{max}}}, \label{eqn:seas_ind}\\
\text{where } D_{j} &= \sum_{i = 1}^{12} p_{i,j} \log_{2}\left(\frac{p_{i,j}}{q}\right).
\end{align}
As before, $p_{i,j}$ is the average case proportion for month $i$ and $q = 1/12$, for $ i = 1, \dots, 12$. $D_{j}$ corresponds to the Kullback–Leibler divergence between the estimated intra-annual distribution and a uniform distribution. Thus, it quantifies how different the monthly proportions are from a uniform distribution over the year. In the context of malaria, $R_{j}$ can be represented by the API at location $j$ and $R_{\text{max}}$ is the maximum API over the region. 
\\
One benefit of using $S_{j}$ is that it separates the timing and amplitude aspects of seasonality. Since high-resolution malaria burden estimates already exist \cite{weiss2019, battle2019, website:map}, the model did not need to estimate the number of cases and could focus exclusively on estimating the monthly proportions at each location. Estimates of the MPI were derived for each location by multiplying the estimated monthly proportions with the mean API. For the Madagascar case study, the 2016 mean \textit{Pf} API estimates were used. 

\subsection*{Deriving seasonality features} \label{sec:season_stats}

Locations were considered as potentially seasonal if their entropy $D_{j} > 0$. When this criterion was satisfied, a rescaled von Mises (RvM) density was fitted via least-squares to the estimated monthly proportions. This is illustrated for an example Malagasy health facility in Supplementary Fig. 6.   
\\
By treating the month in a year as a random variable on a circle, i.e. defining $\theta = \frac{2\pi i}{12}$ where $i = 1, \dots, 12$, the two-component RvM density function can be written as follows:
\begin{align}
f(\theta; s, \omega, \mu_{1}, \kappa_{1}, \mu_{2}, \kappa_{2}) &=  s\left[\omega f_{1}(\theta; \mu_{1}, \kappa_{1}) + (1-\omega) f_{2}(\theta; \mu_{2}, \kappa_{2})\right] \\
\text{where } f_{k}(\theta; \mu_{k}, \kappa_{k}) &= \frac{1}{2\pi I_{0}(\kappa_{k})} \exp\{\kappa_{k}\cos(\theta - \mu_{k})\} \label{eqn:vMcomponent}
\end{align}
is a one-component vM density for $k = 1, 2$ with mean and concentration parameters $\mu_{k}$ and $\kappa_{k}$. Here, $I_{0}$ is the modified Bessel function and $\omega$ is a probability weight. The scale parameter $s>0$ modulates between the continuous density function and monthly proportions over discrete months. 
\\
Instead of identifying characteristics based on the monthly proportion estimates directly, seasonal features were based on fitted vM curves, which further smoothed out the estimates. The benefit of using a circular distribution was the continuity of the curve between the months of December and January. Using vM densities, in particular, is convenient for identifying the peaks of the distribution since these correspond to the mean parameters \cite{pewsey2013}. By comparing the values of the fitted curve, the major and the minor peaks of a bimodal distribution can be identified. Although an arbitrary number of von Mises components can be used, one or two were used because areas with seasonal malaria transmission typically have one or two main seasons \cite{stuckey2014}.
\\
To reduce computational burden, a bimodal distribution was only considered if the error from the fit of a unimodal distribution exceeded a set threshold $\tilde{\epsilon}>0$. For the Madagascar case study, $\tilde{\epsilon}$ was empirically chosen to be $0.0015$. Based on the fitted vM curve, the transmission periods were identified by marking the months where the curve was at or above $\frac{1}{12}$. In this way, the start, end and length of each season could also be estimated. Algorithm \ref{alg:seasonalstats} summarises the procedure used to obtain the seasonality statistics from the monthly proportion realisation curves. 
\\
To quantify the uncertainty associated with the derived statistics, the results from 100 posterior samples of the monthly proportions were summarised. A location was deemed as unimodal or bimodal if more than half of the samples supported that interpretation and the degree of certainty was the proportion of such samples. Based on this majority decision and the corresponding samples, the uncertainty was also assessed in the estimated seasonal characteristics. Circular medians and deviations were used for the start, end and peak of each transmission season \cite{fisher1995}. To interpret the deviations in terms of months, the circular deviations was multiplied by $\frac{12}{2\pi}$.

\begin{algorithm}[tbp]
\caption{Algorithm for deriving seasonality statistics from a monthly proportion realisation curve of location $j$.}
\label{alg:seasonalstats}
\begin{algorithmic}
\Require $p_{i,j} \geq 0$, $\sum\limits_{i = 1}^{12} p_{i,j} = 1$ and error threshold $\tilde{\epsilon} > 0$. 
\State Record $D_{j} = \sum_{i = 1}^{12} p_{i,j} \log_{2}\left(12 p_{i,j}\right)$
\If{$D_{j}>0$}
\State Fit a rescaled, one-component von Mises density to $\{p_{i,j}\}_{i = 1, \dots, 12}$.
\If{the squared error of the fit $> \tilde{\epsilon}$}
\State Fit a rescaled, two-component von Mises density to $\{p_{i,j}\}_{i = 1, \dots, 12}$.
\EndIf
\State Record the major peak months and the minor peak months where applicable.
\State Record the start, end and length of each transmission season based on when the fitted values exceed $\frac{1}{12}$.
\State Label the location as `bimodal' if it has two seasons and `unimodal' otherwise.
\EndIf
\end{algorithmic}
\end{algorithm}

\section*{Data availability} 

The data that support the findings of this study are available from the Programme National de Lutte Contre le Paludisme de Madagascar and the Institut Pasteur de Madagascar (IPM). Health facilities locations were updated and prepared by the SaGEO (Sant\'e et GEOmatique) group at IPM. The source was made available to IPM by the Service d’Information Sanitaire(SIS)/Ministry of Health.

\clearpage

\bibliographystyle{naturemag} 
\bibliography{refs_seasonality}

\section*{Acknowledgments}
We would like to thank the Bill \& Melinda Gates Foundation for funding this research. 

\section*{Author contributions}
M. Nguyen and D.J. Weiss conceived of the presented idea. M. Nguyen designed the analysis, performed the computations and wrote the paper. D.J. Weiss and R.E. Howes advised on and helped shape the research. H.S. Gibson, J. Rozier, S. Keddie, E. Collins and F. Rakotomanana prepared the datasets. All authors provided critical feedback and approved the final manuscript. 

\section*{Additional information}
Additional plots can be found in the Supplementary Information.

\section*{Competing interests}
The authors declare no competing interests.


\end{document}


\maketitle
\pagenumbering{arabic}

\thispagestyle{fancy}

\clearpage

\begin{figure}[tbp]
\centering
\includegraphics[width = 6in, height = 2.4in, trim = 2cm 1cm 2cm 1cm]{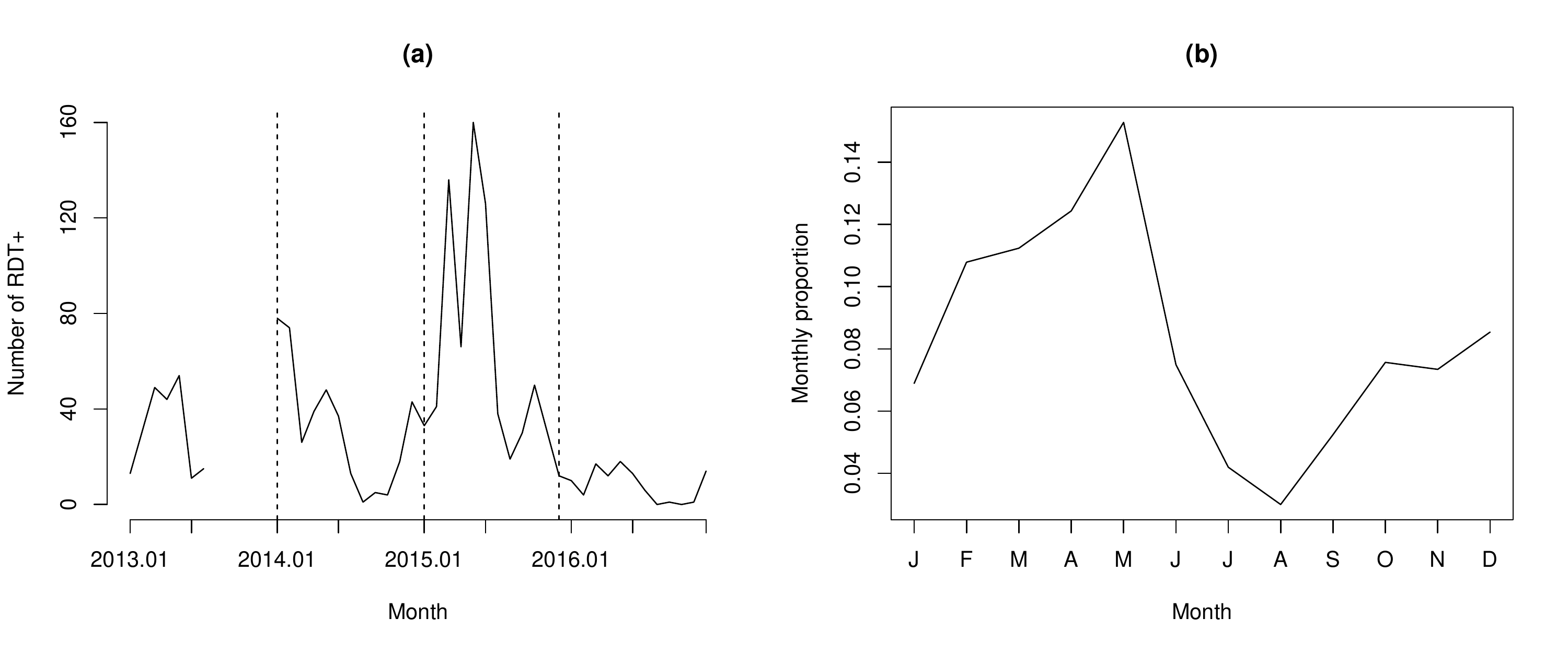}
\caption{(a) Number of positive rapid diagnostic tests (RDTs) for an example Malagasy health centre recorded between 2013 and 2016 and (b) the corresponding monthly proportions computed via dividing the monthly medians by the annual total.}\label{fig:Hf_TS_eg}
\end{figure}

\begin{figure}[tbp]
\centering
\includegraphics[width = 6.4in, height = 3.2in]{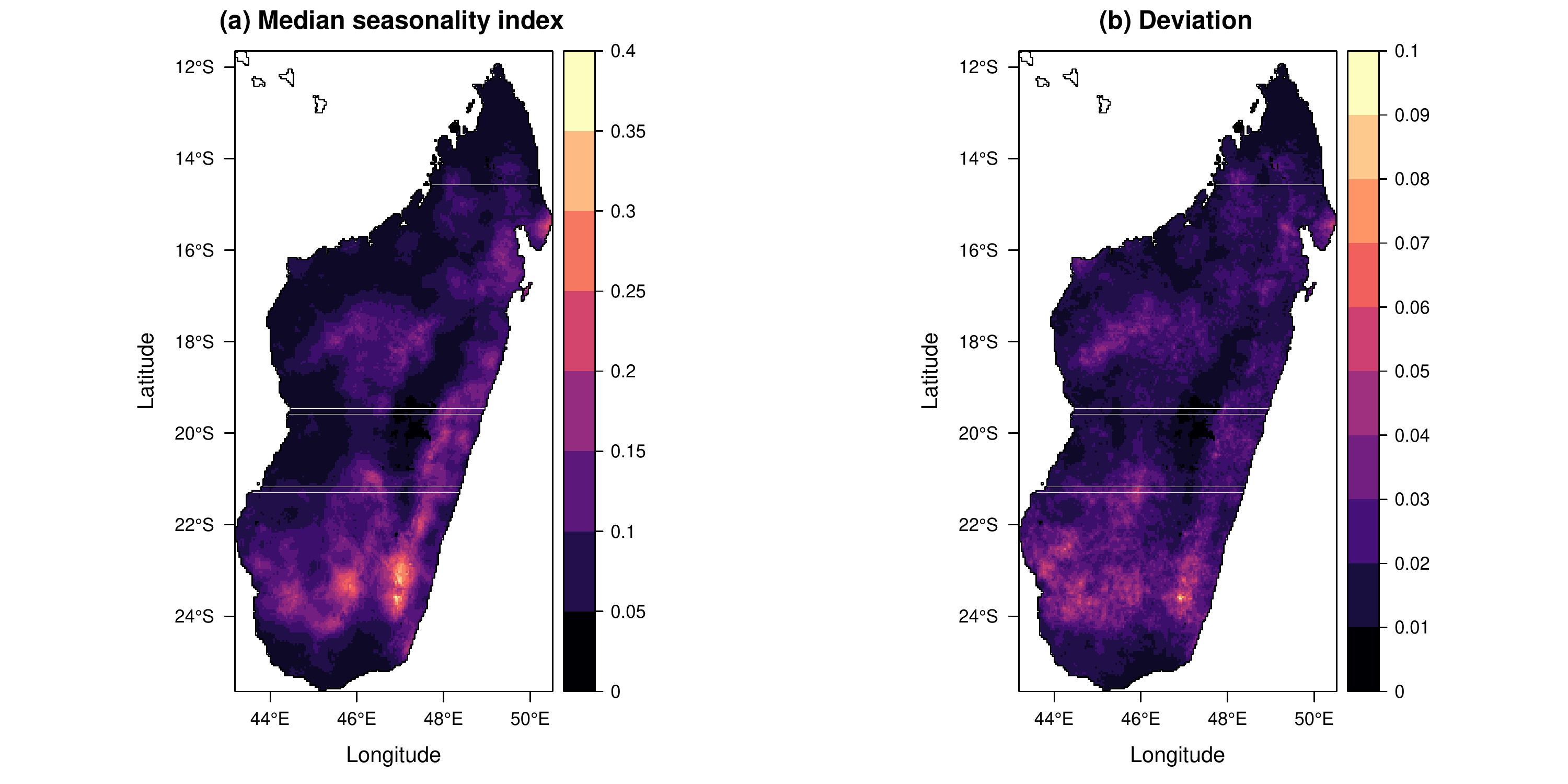}
\caption{(a) Map of the median seasonality index in Madagascar and (b) the associated deviations.}\label{mad_si}
\end{figure}

\begin{figure}[tbp]
\centering
\includegraphics[width = 6in, height = 7.2in, trim = 1cm 0.5cm 1cm 0.5cm]{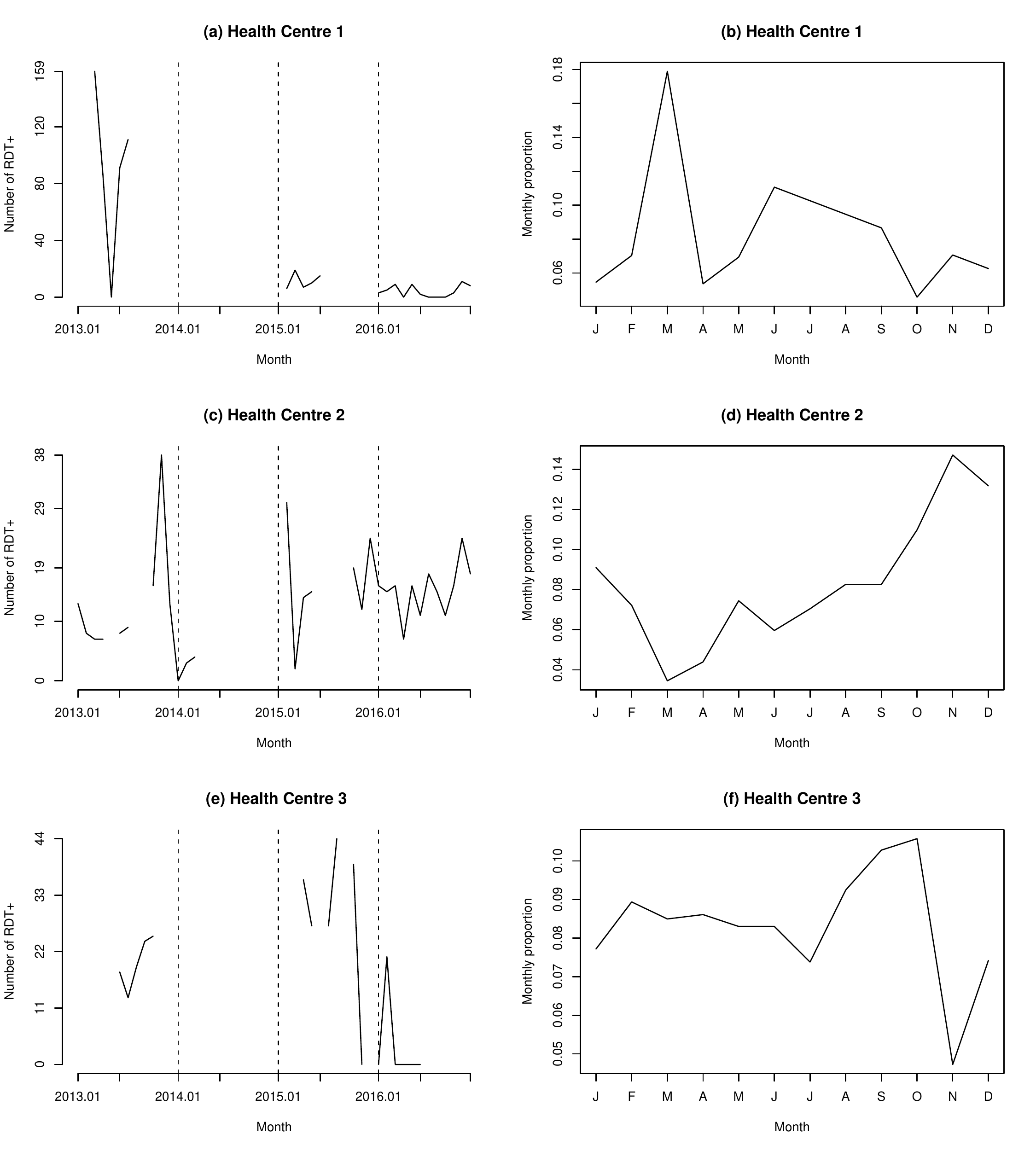}
\caption{Number of positive rapid diagnostic tests (RDTs) recorded between 2013 and 2016 and the corresponding monthly proportions for three example health centres in Melaky.}\label{fig:Melaky_TS_eg}
\end{figure}

\begin{figure}[tbp]
\centering
\includegraphics[width = 4.8in, height = 2.4in]{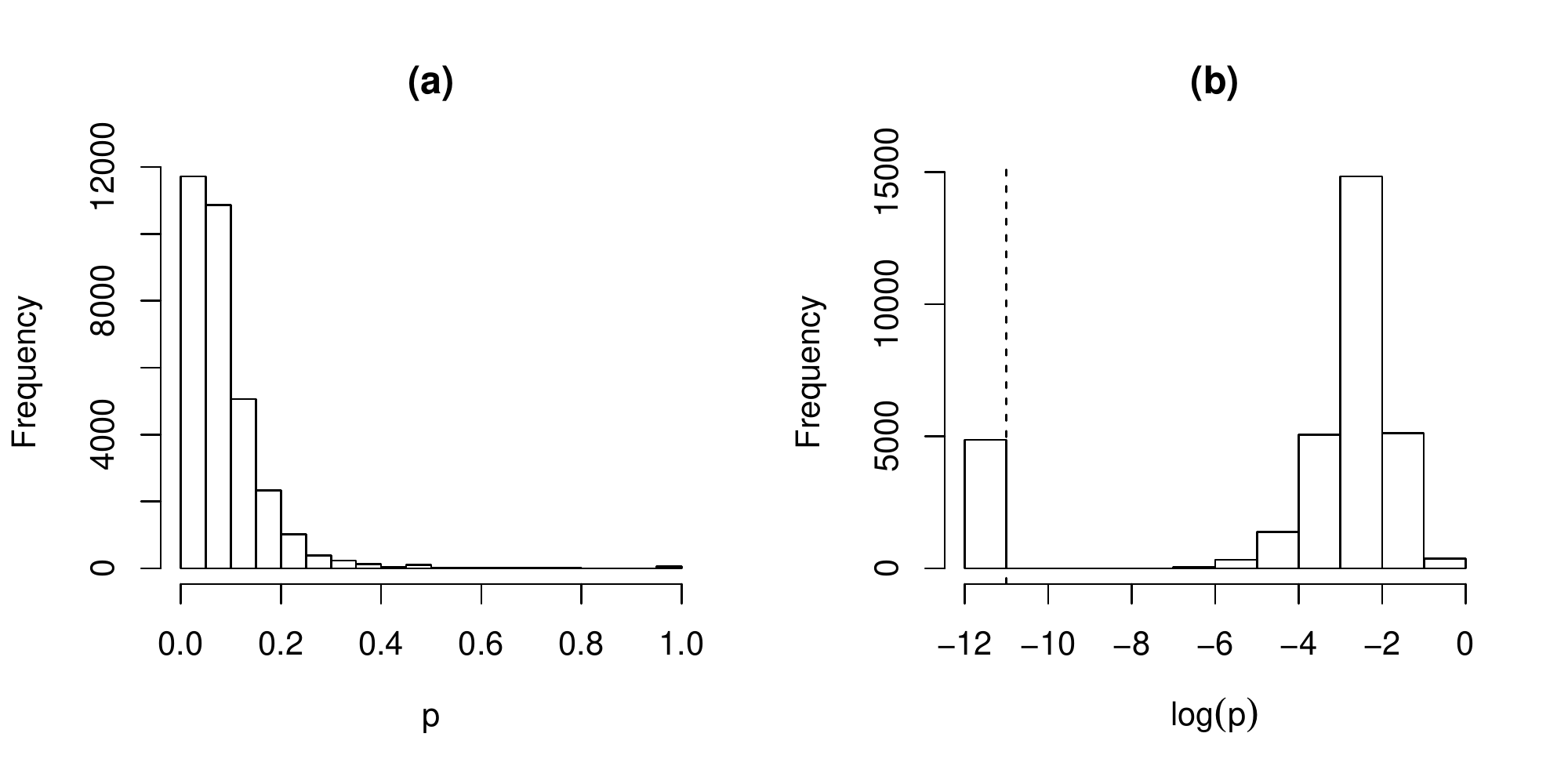}
\caption{(a) Histogram of the empirical monthly proportions ($p_{i, j}$) from the Madagascar health facility data and (b) histogram of the log-transformed monthly proportions ($\log(p_{i, j})$). Since points with $\log(p_{i, j})\leq -11$ are treated as outliers, the dotted vertical line in (b) denotes cut-off.}\label{fig:outlier_hist}
\end{figure}

\begin{figure}[tbp]
\centering
\includegraphics[width = 3.2in, height = 3.2in]{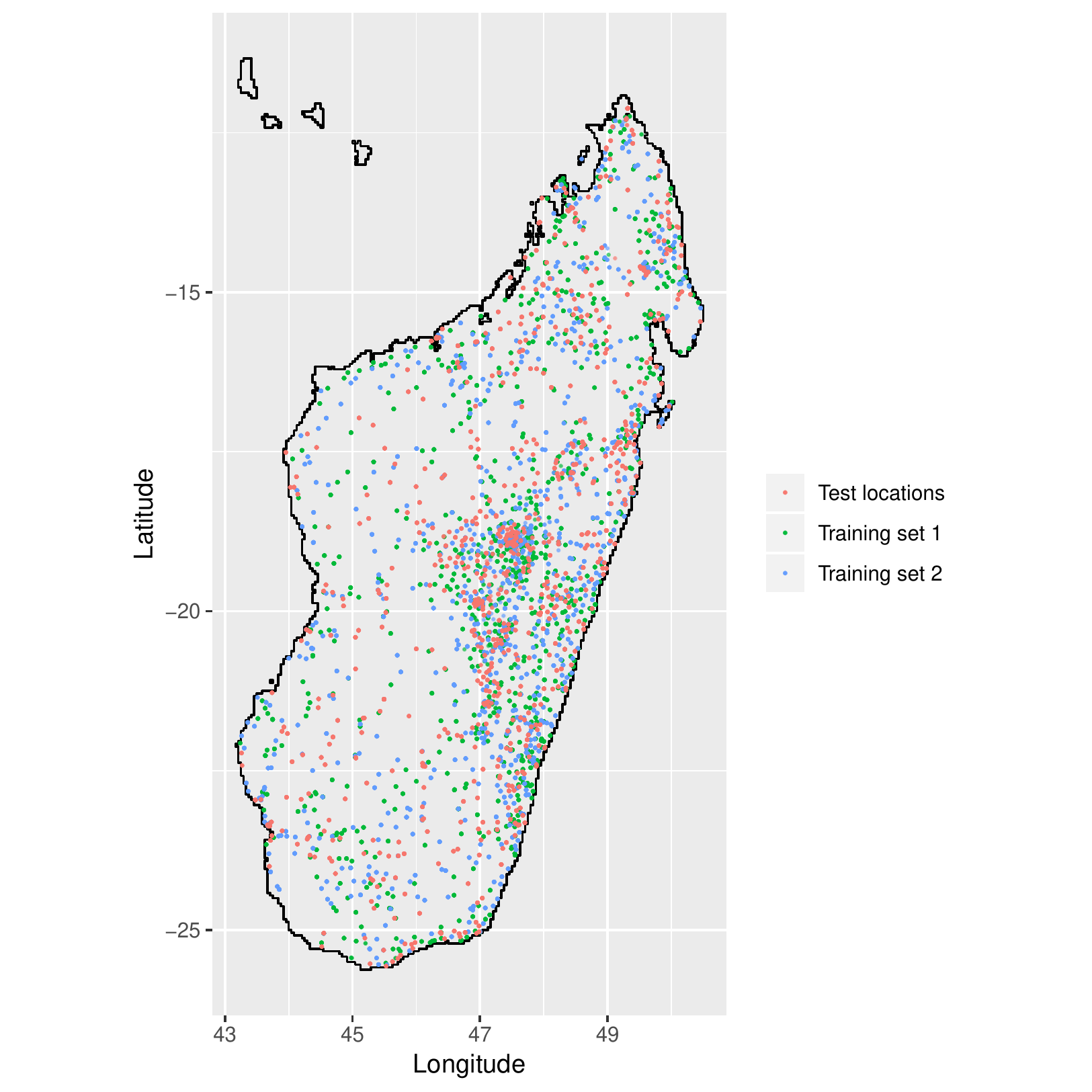}
\caption{Locations of the test and training health facilities for Madagascar.}\label{mad_locations}
\end{figure}

\begin{figure}[tbp]
\centering
\includegraphics[width = 4.2in, height = 2.8in]{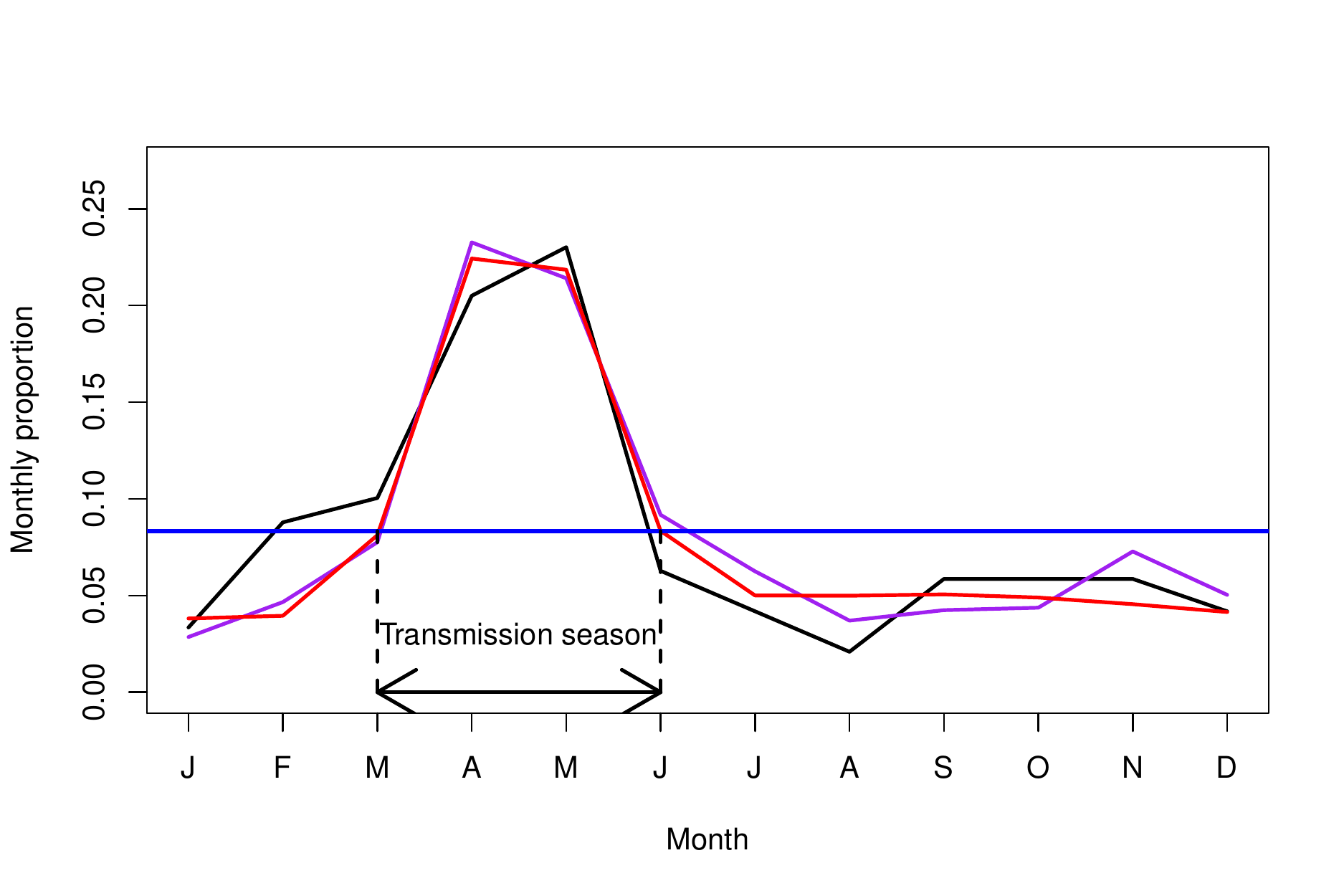}
\caption{An illustration of how the seasonal characteristics can be identified from monthly proportion realisations for an example Malagasy health centre. The empirical proportions are denoted by the black line while one posterior sample is given in purple and its fitted rescaled von Mises density is given in red. The horizontal blue line denotes the $1/12$ threshold and the dotted lines mark out the derived transmission season.}\label{fig:realisation_stat}
\end{figure}


\begin{figure}[tbp]
\centering
\includegraphics[width = 6.4in, height = 3.2in]{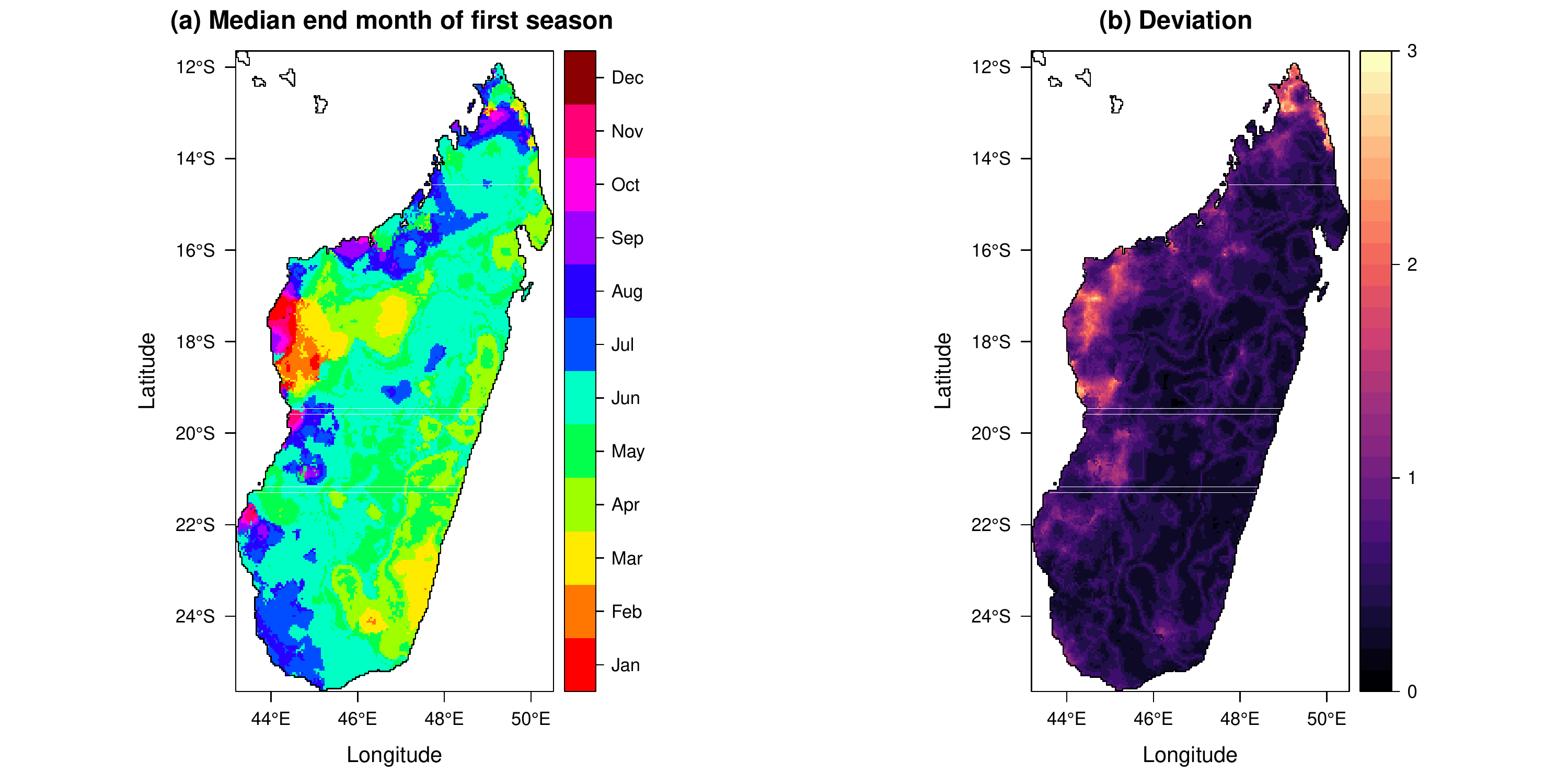}
\caption{(a) Median end months of the first transmission season in Madagascar and (b) the associated deviations.}\label{mad_end1}
\end{figure}

\begin{figure}[tbp]
\centering
\includegraphics[width = 6.4in, height = 3.2in]{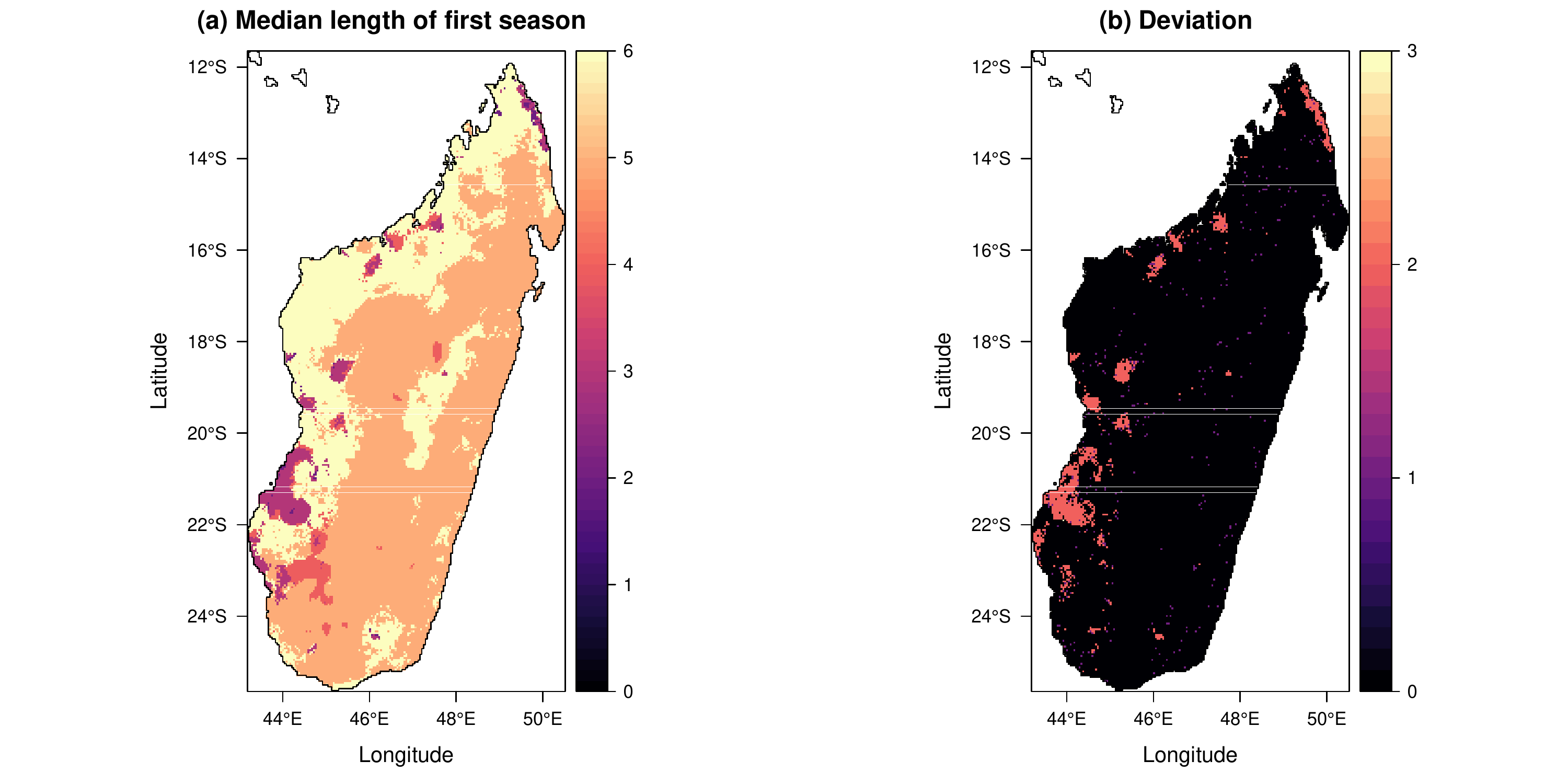}
\caption{(a) Median length (in months) of the first transmission season in Madagascar and (b) the associated deviations.}\label{mad_len1}
\end{figure}

\begin{figure}[tbp]
\centering
\includegraphics[width = 6.4in, height = 3.2in]{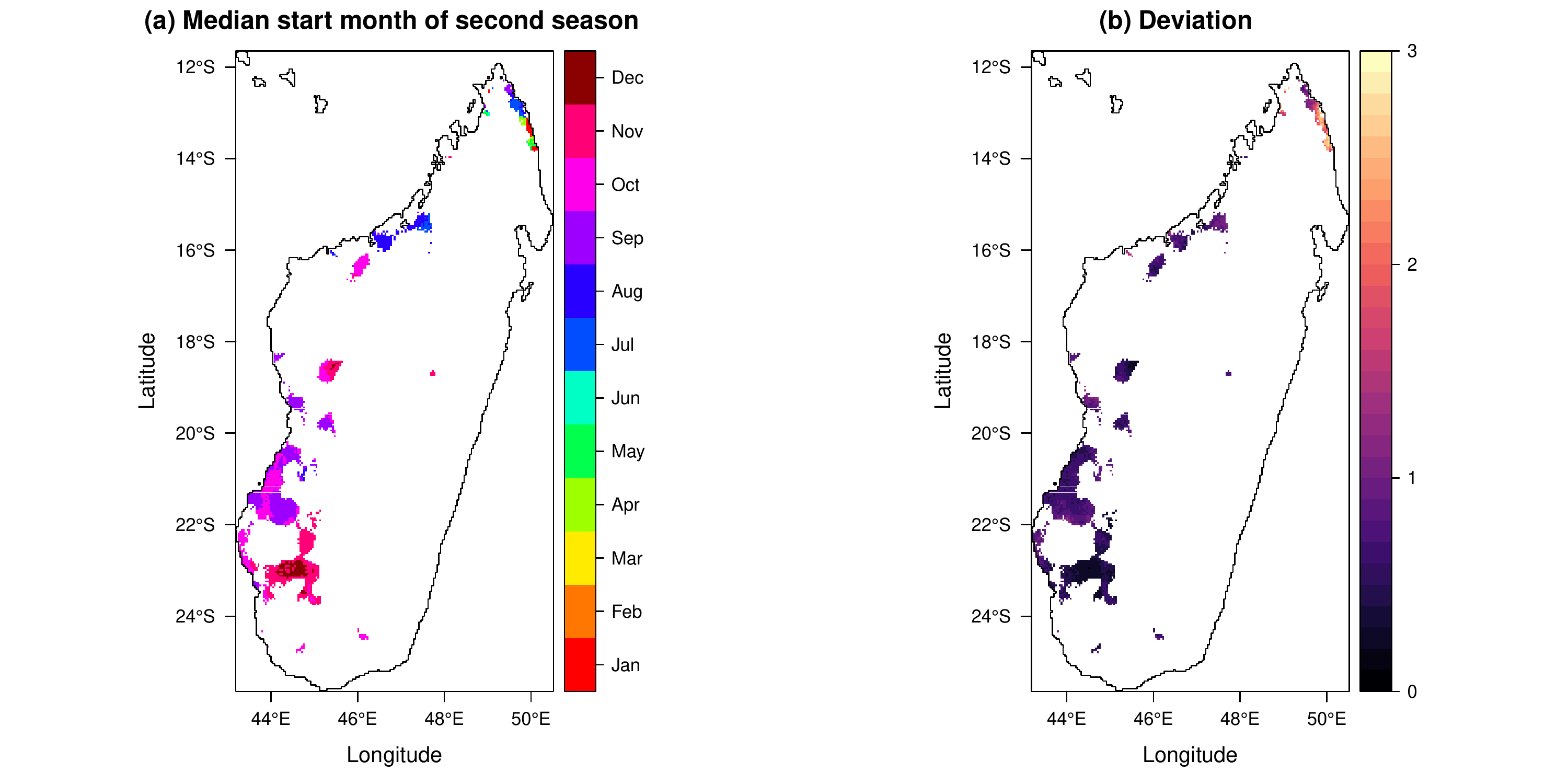}
\caption{(a) Median start months of the second transmission season in Madagascar and (b) the associated deviations. Only the areas deemed as bimodal are coloured.}\label{mad_start2}
\end{figure}

\begin{figure}[tbp]
\centering
\includegraphics[width = 6.4in, height = 3.2in]{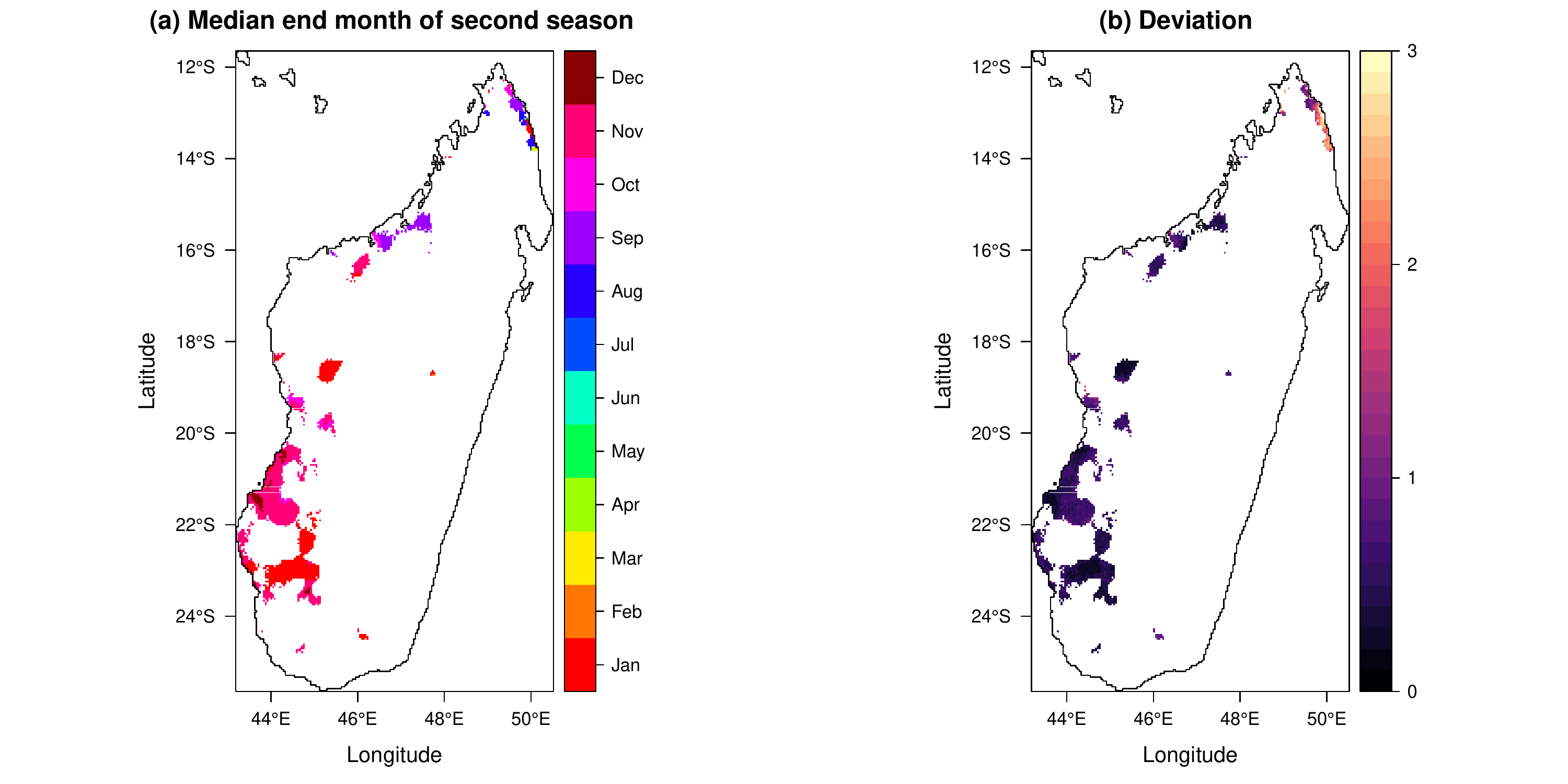}
\caption{(a) Median end months of the second transmission season in Madagascar and (b) the associated deviations. Only the areas deemed as bimodal are coloured.}\label{mad_end2}
\end{figure}

\begin{figure}[tbp]
\centering
\includegraphics[width = 6.4in, height = 3.2in]{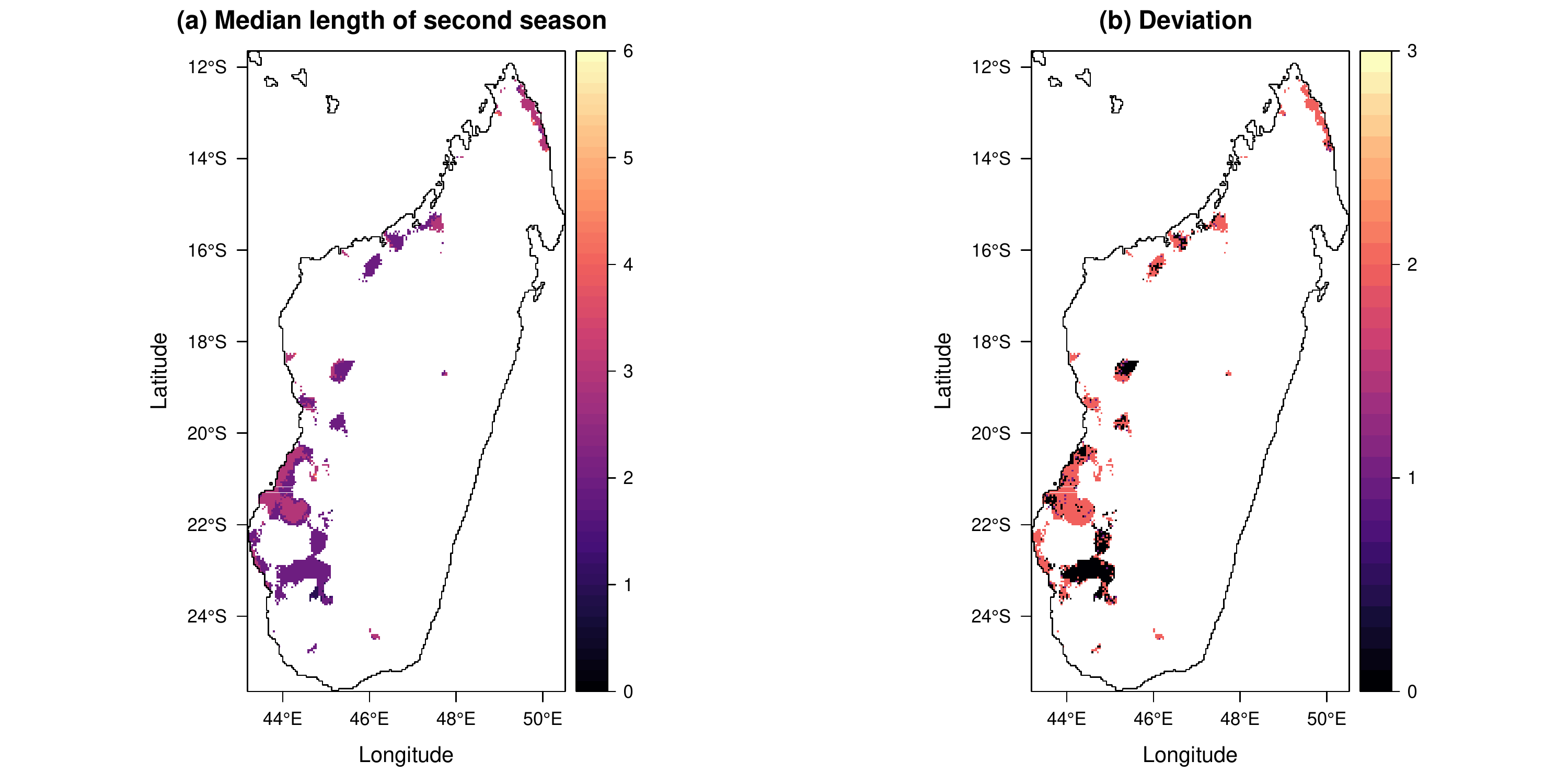}
\caption{(a) Median length (in months) of the second transmission season in Madagascar and (b) the associated deviations. Only the areas deemed as bimodal are coloured.}\label{mad_len2}
\end{figure}

\begin{figure}[tbp]
\centering
\includegraphics[width = 6.4in, height = 3.2in]{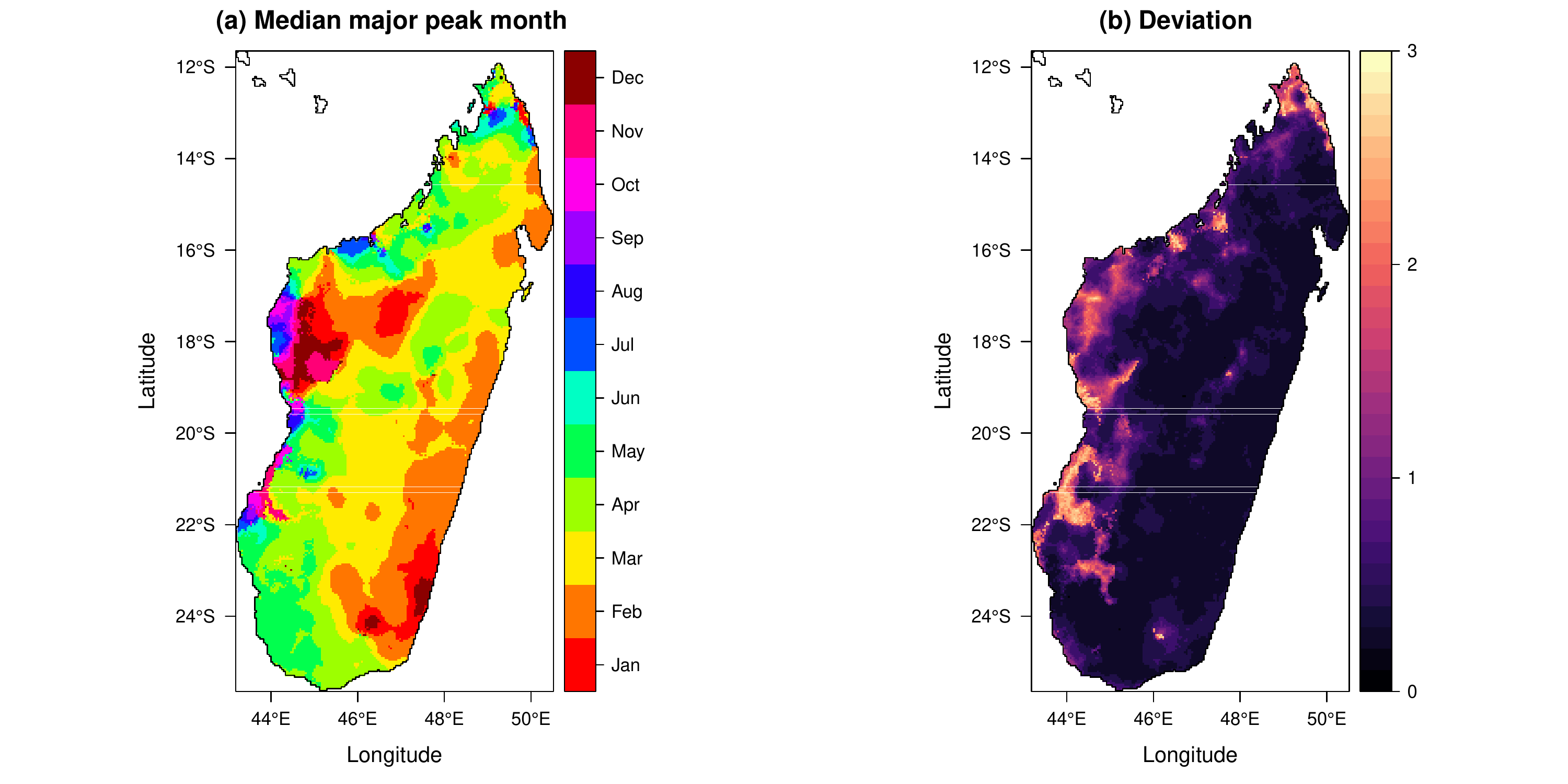}
\caption{(a) Median major peak months in Madagascar and (b) the associated deviations. Note that this is the single peak for unimodal locations.}\label{mad_majpeak}
\end{figure}

\begin{figure}[tbp]
\centering
\includegraphics[width = 6.4in, height = 3.2in]{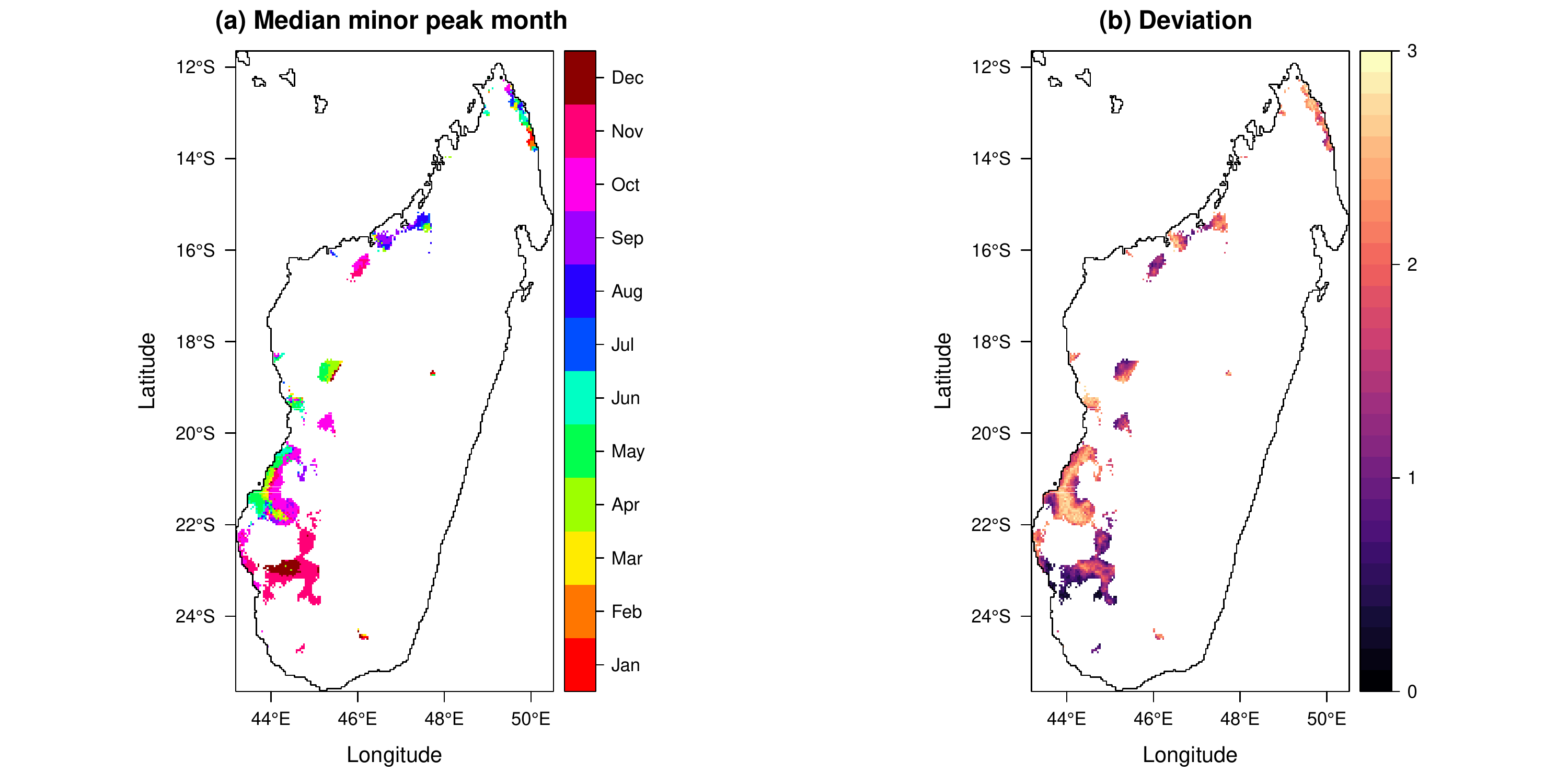}
\caption{(a) Median minor peak months in Madagascar and (b) the associated deviations. This is only applicable for areas deemed as bimodal.}\label{mad_minpeak}
\end{figure}

